\documentclass{amsart}
\usepackage{graphicx}
\usepackage{color}
\usepackage{amssymb}
\usepackage{amsthm}
\usepackage{pxfonts}
\usepackage{latexsym}
\usepackage{subfigure}
\usepackage{url}
\usepackage{algorithmicx}
\usepackage{multirow}

\newcommand{\ud}{\textup{d}}

\newcommand{\NN}{\mathbb{N}}
\newcommand{\ZZ}{\mathbb{Z}}

\newcommand{\RR}{\mathbb{R}}

\newcommand{\II}{\textnormal{\tiny \textsc{II}}}
\newcommand{\QRS}{\textnormal{\tiny \textsc{QRS}}}
\newcommand{\beat}{\textnormal{\tiny \textsc{beat}}}

\begin{document}
\title[Heart beat classification from single-lead ECG]{Heart beat classification from single-lead ECG using the Synchrosqueezing Transform}

\author{Christophe~L~Herry, Martin~Frasch, Andrew~JE~Seely and Hau-tieng~Wu}
\address{Ottawa Hospital Research Institute, Ottawa, Ontario, Canada email: (cherry@ohri.ca)}
\address{University of Washington, Seattle, Washington, USA}
\address{The Ottawa Hospital, Ottawa, Ontario, Canada}
\address{Department of Mathematics, University of Toronto, Toronto, Ontario, Canada and Mathematics Division, National Center for Theoretical Sciences, Taipei, Taiwan.}

\maketitle

\begin{abstract}
The processing of ECG signal provides a wealth of information on cardiac function and overall cardiovascular health. While multi-lead ECG recordings are often necessary for a proper assessment of cardiac rhythms, they are not always available or practical, for example in fetal ECG applications. Moreover, a wide range of small non-obtrusive single-lead ECG ambulatory monitoring devices are now available, from which heart rate variability (HRV) and other health-related metrics are derived. Proper beat detection and classification of abnormal rhythms is important for reliable HRV assessment and can be challenging in single-lead ECG monitoring devices.
In this manuscript, we modelled the heart rate signal as an adaptive non-harmonic model and used the newly developed synchrosqueezing transform (SST) to characterize ECG patterns. We show how the proposed model can be used to enhance heart beat detection and classification between normal and abnormal rhythms. In particular, using the Massachusetts Institute of Technology-Beth Israel Hospital (MIT-BIH) arrhythmia database and the Association for the Advancement of Medical Instrumentation (AAMI) beat classes, we trained and validated a support vector machine (SVM) classifier on a portion of the annotated beat database using the SST-derived instantaneous phase, the R-peak amplitudes and R-peak to R-peak interval durations, based on a single ECG lead. We obtained sentivities and positive predictive values comparable to other published algorithms using multiple leads and many more features.   
\end{abstract}

\noindent{\it Keywords\/}: R-peak detection, synchrosqueezing transform, instantaneous phase, instantaneous frequency, heartbeat classification

\section{Introduction}
The recording of the electrical activity of the heart by means of an Electrocardiogram (ECG) is ubiquitous in healthcare settings, as well as in daily environments. The processing of ECG signal provides a wealth of information on cardiac function and overall cardiovascular health; it also offers a unique non-invasive window into pathological and dysfunctional states of the human body, via e.g. heart rate variability (HRV) assessment. While multi-lead ECG recordings are necessary for a proper assessment of cardiac rhythms, more often only one-lead recordings are available for analysis, thereby complicating the delineation of fiducial points (P,Q,R,S,T) and the derivation of their characteristic amplitudes and timings. This is especially relevant in fetal heart rate monitoring, where no satisfying standard in fetal ECG derivation has been accomplished. Many ambulatory monitoring systems and modern mobile devices are also single lead ECG and typically have uncommon lead configuration due to the placement of the device.

One important aspect of ECG processing, in addition to the common pre-processing steps to remove noise, baseline wandering, muscular artefacts is the extraction of beats and delineation of QRS complexes. Numerous algorithms have been proposed to detect R-peaks, the first stage in ECG delineation, many of them providing excellent results in a variety of noisy or arrhythmic scenarios. When tested on publicly available databases such as the MIT-BIH arrhythmia database, the top tier algorithms typically achieve above 99\% Sensitivity (Se) and Positive Predictive Value (+P). Restricting ourselves to single-lead recordings, examples of good performing algorithms include the classic algorithm proposed by Pan and Tompkins \cite{Pan_Tompkins:1985} and later improved by Hamilton and Tompkins \cite{Hamilton_Tompkins:1986}, algorithms based on differentiation \cite{Yeh2008}, wavelet-based detection \cite{Martinez2004}, thresholding of Shannon energy \cite{Kathirvel2011}, Empirical Mode Decomposition \cite{Li2014} or combination of moving averages \cite{Elgendi2013}. While perhaps non-critical in common monitoring scenarios, the number of missed or false beats can adversely affect HRV calculations by artificially increasing/decreasing the time intervals between R-peaks (RRI) and hence skew the estimates of HRV.

Another important aspect of ECG processing is the identification of cardiac arrhythmias, usually differentiated as life-threatening or non life-threatening arrhythmias. Life-threatening arrhythmias such as ventricular fibrillation or tachycardia have been thoroughly investigated in the past \cite{Aramendi2010,Clayton1993}. Non emergency arrhythmias are comprised of a large number of abnormal beat types that can occur sporadically. Automatic heartbeat classification is therefore desirable for an efficient monitoring of patients. Numerous algorithms have been proposed to classify non life-threatening arrhythmic  beats \cite{Dechazal2004,llamedo2011heartbeat,ye2012heartbeat,de2012weighted}, using ECG morphological features, time and frequency features \cite{Dechazal2004,ye2012heartbeat,llamedo2011heartbeat} or higher-order statistics \cite{osowski2004support}. 
 
In this manuscript we aim to enhance the R-peak detection, quality assessment and detection of the ECG arrhythmic patterns using the recently developed synchrosqueezing transform (SST) \cite{Daubechies_Lu_Wu:2011,Chen_Cheng_Wu:2014}, which can help study time-varying oscillatory patterns characteristic of heart rhythms. For the ECG beat type classification, we focus our analysis on single-lead ECG as we anticipate the greatest gain in applications such as fetal heart rate monitoring and simple ambulatory monitoring devices.

The paper is organized in the following way. In Section \ref{Section:synchrosqueezing}, we summarize the main algorithm, the synchrosqueezing transform; in Section \ref{Section:RpeakAlgo}, the R-peak detection algorithm based on \cite{Elgendi2013} is discussed; in Section \ref{Section:OurAlgorithm}, we proposed our algorithm by integrating SST and R-peak algorithm; in Section \ref{Section:Result}, the analysis result applied to the standard MIT-BIH benchmark is showed; the discussion and future work are in Section \ref{Section:Discussion}.

\section{Adaptive non-harmonic model and Synchrosqueezing transform}\label{Section:synchrosqueezing}
The main signal processing tool we propose to boost existing R-peak detection algorithms is the newly developed, mathematically rigorously proved, and adaptive time-frequency (TF) analysis referred to as the synchrosqueezing transform (SST) \cite{Daubechies_Lu_Wu:2011,Chen_Cheng_Wu:2014}. The main motivation of SST is to study the ubiquitous oscillatory signals, in particular when the oscillatory pattern is not fixed from time to time\footnote{Sometimes this kind of time-varying pattern is referred to as ``nonlinear'' or ``non stationary''. To avoid possible confusions with rigorously defined mathematical terminologies, we do not use these terms.}. Precisely, it is well known that any specified physiological system intimately interacts with other physiological systems in complex ways; for example, chemical set points, metabolic demand, neural dynamics, etc, play a role in not only the electrophysiological conducting pathway inside the heart, but also the respiration patterns \cite{Benchetrit:2000}, which further contributes as one source of the heart rate variability \cite{Malik_Camm:1995}. Several other examples can be found in circadian, cortical rhythms, etc \cite{Takeda_Maemura:2011,Wang:2010,Baudin_Wu_Bordessoule_Beck_Jouvet_Frasch_Emeriaud:2014,Seely_Bravi_Herry_Green:2014}. This complex behavior is reflected in the oscillatory signal we observe -- the oscillation might be fast in the beginning and slow later. These complicated time varying fast-slow and large-small oscillatory behaviors will be modeled by an {\it adaptive (non-) harmonic model}.

Extracting such time-varying features has been attracting many researchers' attention in the past decades. Short time Fourier transform (STFT), continuous wavelet transform (CWT) and other time-frequency (TF) analysis techniques \cite{Flandrin:1999,Daubechies:1992} have been widely applied in the past to extract such features by studying the signal's time frequency representation. However, due to the uncertainty principle, the ambiguity in the classical TF representation determined by the STFT and the CWT is inevitable \cite{Daubechies:1992,Flandrin:1999}. For practical purposes, a technique to sharpen the TF representation was proposed in \cite{Kodera_Gendrin_Villedary:1978}, which is known as the reassignment technique. The reassignment technique was later improved in \cite{Auger_Flandrin:1995,Chassande-Mottin_Auger_Flandrin:2003}. Among reallocation techniques, SST is a special one. It enjoys several different properties, for example, oscillatory component reconstruction \cite{Daubechies_Lu_Wu:2011,Chen_Cheng_Wu:2014} and real time implementation \cite{Chui_Lin_Wu:2014}, and it has been applied to several different clinical problems to extract the physiological dynamics hidden inside different observed oscillatory signals. For example, the sleep dynamics can be estimated by analyzing the respiratory signal \cite{Wu:2013,Chen_Cheng_Wu:2014,Wu_Talmon_Lo:2014}, the subcortical activity during general anesthesia can be observed by analyzing the ECG's R peak to R peak interval time series \cite{Wu_Chan_Lin_Yeh:2013,Lin_Wu_Tsao_Yien_Hseu:2013}, the ventilator weaning outcome can be predicted by analyzing the respiratory flow and ECG signals \cite{Wu_Hseu_Bien_Kou_Daubechies:2013}, etc.

We conjecture that based on the adaptive (non-)harmonic model and SST, finding these time-varying features associated with the oscillatory signal might help us to better understand the system, in particular the heart beating pattern, and hence increase the R peak detection accuracy.

\subsection{Adaptive non-harmonic model for the ECG signals}
Here we recall the adaptive non-harmonic model, and discuss the significance of the ECG signal to motivate a generalized model. For interested readers, we would refer to \cite{Daubechies_Lu_Wu:2011,Wu:2011,Wu:2013,Chen_Cheng_Wu:2014}. The {\em adaptive (non-)harmonic model}, proposed in \cite{Wu:2013}, can be viewed as a phenomenological model describing the observed oscillatory signals, which is a generalization of the well-known harmonic model in Fourier analysis. To simplify the discussion and focus on the R peak detection problem of interest, we discuss the one oscillatory component case and its generalization. For the model consisting of multiple oscillatory components with a trend, we refer to \cite{Daubechies_Lu_Wu:2011,Wu:2013,Chen_Cheng_Wu:2014}. 
\begin{equation}\label{eq:resp_model}
Y(t)=A(t)s(\phi(t))+\Phi(t),
\end{equation}
where 
\begin{enumerate}
\item[(I)] $A\in C^1(\RR)\cap L^\infty(\RR)$, $\phi\in C^2(\RR)$, $\inf_{t\in\RR} A(t)>0$, $\inf_{t\in\RR}\phi'(t)>0$, $|A'(t)|\leq \epsilon \phi'(t)$ and $|\phi''(t)|\leq \epsilon\phi'(t)$, for all $t$, and $\epsilon$ is a small parameter;
\item[(II)] $s:[0,1]\to \RR$ is $C^{1,\alpha}$, where $\alpha>1/2$, and $1$-periodic function with unit $L^2$ norm, $|\hat{s}(k)|\leq \delta |\hat{s}(1)|$ for all $k\neq 1$, where $\delta\geq 0$ is a small parameter, and $\sum_{n>D}|n\hat{s}(n)|\leq \theta$ for some small parameter $\theta\geq 0$ and $D\in \NN$. 
\item[(III)] $\Phi(t)$ is a stationary random process or ``almost'' stationary random process \cite{Chen_Cheng_Wu:2014} independent of $A(t)s(\phi(t))$, which is introduced to model the measurement error. A common example of $\Phi$ is a Gaussian white noise.
\end{enumerate}
The condition (I) essentially quantifies how a signal oscillates from time to time -- we call $A(t)$ the {\it amplitude modulation function}, $\phi$ the {\it phase function} and $\phi'(t)$ the {\it instantaneous frequency}. Clearly, the larger the $\phi'(t)$ is, the faster the signal oscillates. The condition (II) is critical for our application. It is clear that although an ECG signal is oscillatory, the cycle is far from a cosine function. Thus, we take a $1$-periodic function to approximate this non-cosine oscillation. We call $s$ the {\it wave shape function}. When the wave shape function is $\cos(t)$, we call the model {\em adaptive harmonic model}. For more discussion about wave shape function, we refer the reader to \cite{Wu:2013}. The condition (III) is inevitable since noise is everywhere during the measurement.

With the above adaptive non-harmonic model, we could now model the ECG signal. It has been widely accepted that recorded ECG signals are the projection of the representative dipole current of the electrophysiological cardiac activity on different directions \cite{Keener:1998}; that is, the representative dipole current, denoted as $d(t)\in\RR^3$, where $t\in\RR$, is projected on $v_\ell\in \RR^3$, where $\ell$ is the index of the ECG channel. Denoted the $\ell$-th ECG channel as $E_\ell(t)=v_\ell^Td(t)$\footnote{In general, $v_\ell$ should change according to time due to the respiratory activity and other physical movement. To simplify the discussion here, we do not take these facts into account}. It is clear that $E_\ell(t)$ is oscillatory but the oscillation cannot be modeled by a cosine function.
At first glance, we could model the ECG signal by the adaptive non-harmonic model (\ref{eq:resp_model}), and have
\begin{equation}
E_\II(t)=A_\II(t)s_\II(\phi(t))\label{eq:resp_model1}\approx\sum_{l=1}^DA_\II(t)\hat{s}_{\II,l}\cos(2\pi l\phi(t)+\psi_{\II,l}),\nonumber
\end{equation}
where $\psi_{\II,l}\in [0,2\pi)$ and $\hat{s}_{\II,l}$ is the Fourier series coefficients of the wave shape function $s_\II$ modeling the oscillation inside the ECG signal. Note that $\psi_{\II,l}$, $l=1,\ldots,D$ depends only on $s_\II$. Therefore, we can view the lead II ECG signal as the single component adaptive harmonic model $A(t)\hat{s}_\II(1)\cos(2\pi\phi(t)+\psi_{\II,1})$ plus its multiples with proper global phase shifts. 
As useful as it is to describe the oscillatory pattern, however, this adaptive non-harmonic model is not an exact model of the ECG signal and further considerations are needed. Below we take the lead II ECG signal as an example. The same argument holds for other ECG leads.

The main reason why (\ref{eq:resp_model}) is not an exact model is that the oscillatory pattern of the ECG signal is clearly not fixed from one to another. 
Take the QT interval as an example. The QT interval depends on the heart rate and a prolonged interval might indicate pathological status. The influence of heart rate on the QT interval interpretation and the relationship between QT interval and heart rate have been broadly studied. The common models include Bazett's formula (QT interval is proportional to the square root of RR interval) \cite{Bazett:1920} and Fridericia's formula (QT interval is proportional to the cubic root of RR interval) \cite{Fridericia:1920}. Although there are several debates on how to improve these models, it is generally accepted that the relationship between the QT interval and the heart rate cannot be linear \cite{Fossa_Zhou:2010}. On the other hand, note that in the adaptive non-harmonic model (\ref{eq:resp_model}), the QT interval is linearly related to the RR interval. Indeed, suppose the $k$-th R peak happens at time $t_k$, by the mean value theorem we have the following relationship for time $t\in[t_k,t_{k+1}]$ 
\begin{equation}
s_\II(\phi(t))\,=\sum_{k\in\ZZ}s_\II(\phi(t_k)+(t-t_k)\phi'(\tilde{t}))\chi_{[t_k,t_{k+1})}\,=\sum_{k\in\ZZ}s_\II\left(\frac{t-t_k}{1/\phi'(\tilde{t}_k)}\right)\chi_{[t_k,t_{k+1})},\label{Model:LinearRelationship}
\end{equation}
where $\tilde{t}_k\in[t_k,t_{k+1}]$, $\chi_{[t_k,t_{k+1})}$ is a indicator function defined on $[t_k,t_{k+1})$, and the second equality holds since $\phi(t_k)=k$ and $s$ is $1$-periodic. While the RR interval between the $k$-th and the $(k+1)$-th R peaks is proportional to $1/\phi'(\tilde{t})$ up to order $O(\epsilon)$ by the assumption of $\phi$, we know that the wave shape function is approximately linearly dilated according to $1/\phi'(\tilde{t})$, and hence the claim.
Thus, the adaptive non-harmonic model is not an exact but an approximate model in the sense that the QT interval is linearly related to the heart rate.

As a result, we should consider a generalization of the adaptive non-harmonic model (\ref{eq:resp_model}) to model the clean lead II ECG signal, denoted as $E_\II(t)$:
\begin{equation}
E_\II(t)=\sum_{l=1}^DA_\II(t)c_{\II,l}(t)\cos(2\pi (l\phi(t)+\psi_{\II,l}(t)))\label{eq:resp_model2}=\sum_{l=1}^DB_{\II,l}(t)\cos(2\pi \varphi_l(t)),
\end{equation}
where $\psi_{\II,l}(t)\in C^2(\RR)$ satisfying $|\psi'_{\II,l}(t)|\leq \epsilon \phi'(t)$ and $|\psi''_{\II,l}(t)|\leq \epsilon \phi'(t)$, $c_{\II,l}(t)\in C^1(\RR)$ is a slowly varying function satisfying $|c_{\II,l}'(t)|\leq \epsilon\phi'(t)$, $B_{\II,l}(t):=A_\II(t)c_{\II,l}(t)$ and $\varphi_l=l\phi(t)+\psi_{\II,l}(t)$.
Clearly, when $c_{\II,l}(t)=\hat{s}_{\II,l}$ are constants and $\psi_{\II,l}(t)$ are constants, (\ref{eq:resp_model2}) is reduced to the model (\ref{eq:resp_model1}). Thus, we could either view the model (\ref{eq:resp_model2}) as an adaptive non-harmonic model with one oscillatory component and the {\it time-varying wave shape function}. We would call $\varphi_1$ the {\it fundamental phase function} of the lead II ECG signal, and call $\varphi_\ell$ the {\it $\ell$-th multiple} of the fundamental phase function, although $\varphi_\ell$ might not be proportional to $\varphi_1$. Clearly, this is a direct generalization of the notion ``integral multiple'' in the Fourier analysis. 
Based on the expansion, we could also view (\ref{eq:resp_model2}) as an adaptive harmonic model with $D$ oscillatory components. Note that in (\ref{eq:resp_model1}), we could define the $k$-th oscillatory cycle is the function on the period $[t_k,t_{k+1}]$, where $\phi(t_k)=k$. In (\ref{eq:resp_model2}), the situation is clearly different but we could consider the same definition -- the $k$-th oscillatory cycle is the function on the period $[t'_k,t'_{k+1}]$, where $\varphi_1(t'_k)=k$.  

Finally, we comment that since the cardiac axis varies from one subject to another due to the variable body torso constitution, breathing pattern, and so on, even if we focus on the lead II ECG signal, we cannot find a precise wave shape function for every case at every moment. However, while these variations do not dominate the main feature of interest -- the R peaks, we could assume the existence of a wave shape function which approximates well the lead II signal, in particular when the subject is normal. 

\subsection{Synchrosqueezing transform (SST)}

The SST algorithm is composed of 2 main steps. First, we evaluate a preferred time-frequency analysis, for example, STFT or CWT. Second, evaluate the reassignment rule and squeeze the coefficients in the time-frequency representation. Here we take CWT as an example. The STFT version can be found in \cite{Wu:2011,Oberlin:2015}. Take an observed signal $Y$ which satisfies the adaptive harmonic model. Take a mother wavelet $\psi$ for CWT so that the support of $\widehat{\psi}$ is inside $[1-\Delta,1+\Delta]$, where $0<\Delta<1$. Then calculate the CWT $W_Y(a,b)$ of $Y(t)$, where $a>0$ is the scale and $b\in\RR$ is the time. Next, calculate the reassignment rule, which is denoted as $\omega_Y$: 
\begin{equation}\label{alogithm:sst:ressigment}
\omega_Y(a,b):=\left\{\begin{array}{ll}
\displaystyle\frac{-i\partial_bW_Y(a,b)}{2\pi W_Y(a,b)} & \mbox{when}\quad |W_Y(a,b)|\neq 0;\\
\infty&\mbox{when}\quad |W_Y(a,b)|=0.
\end{array}\right.
\end{equation}
Lastly, the SST of $Y(t)$ is defined by re-allocating the coefficients of $W_Y(a,b)$ according to the reallocation rule $\omega_f(a,b)$:
\begin{equation}\label{alogithm:sst:formula}
S^{\Gamma}_Y(b,\xi):=\lim_{\alpha>0,\alpha\to 0}\hspace{-10pt}\int\limits_{\{(a,b):~|W_Y(a,b)|\geq {\Gamma}\}}\hspace{-20pt}\frac{1}{\alpha}h\left(\frac{|\omega_Y(a,b)-\xi|}{\alpha}\right)W_Y(a,b)a^{-3/2}\ud a
\end{equation}
where $\Gamma>0$ is the threshold chosen by the user and $h$ is a kernel function which is smooth enough and decays fast enough so that $\frac{1}{\alpha}h\left(\frac{\cdot}{\alpha}\right)$ converges waking to the Dirac delta function as $\alpha\to 0$. Intuitively, at each time point $b$, we collect all CWT coefficients indicating the existence of oscillatory components with frequency $\xi$ and put them in the $(b,\xi)$ slot. As is shown in \cite{Daubechies_Lu_Wu:2011,Chen_Cheng_Wu:2014}, $S^{\Gamma}_Y(b,\xi)$ will only have dominant values around $\phi'(b)$. This property allows us an accurate estimate of the instantaneous frequency $\phi'$ by, for example, a curve extraction technique. The estimated $\phi'$ is denoted by $\widetilde{\phi}'$.
With the estimated $\phi'$, we can reconstruct the component and estimate the amplitude modulation $A(t)$ and the phase function $\phi(t)$ by $\mathfrak{Re}\widetilde{R}(t)$, where $\mathfrak{Re}$ means taking the real part,
\begin{equation}
\widetilde{R}(t):=\mathcal{R}_\psi^{-1}\int_{\{\xi:~|\phi'(t)-\xi|\leq \epsilon^{1/3}\}}S_Y^\Gamma(t,\xi)\ud \xi , \label{alogithm:sst:reconstruction}
\end{equation}
and $\mathcal{R}_\psi=\int \frac{\widehat{\psi}(\zeta)}{\zeta}\ud \zeta$. Based on the Theorem in \cite{Daubechies_Lu_Wu:2011,Wu:2013,Chen_Cheng_Wu:2014}, the estimator of $A(t)$ is defined as
$\widetilde{A}(t):=|\widetilde{R}(t)|$, and hence an estimator for $\phi(t)$, denoted as $\widetilde{\phi}(t)$, can be obtained by unwrapping the phase of the complex-valued signal $\widetilde{R}(t)/\widetilde{A}(t)$. 

Here we summarize the theoretical results of SST relevant to our work \cite{Daubechies_Lu_Wu:2011,Chen_Cheng_Wu:2014}.
\begin{itemize}
\item[(P1)] SST is robust to the several different kinds of noise, which might be slightly non-stationary. Thus we are able to accurately estimate the dynamics inside the signal \cite{Chen_Cheng_Wu:2014};
\item[(P2)] Since SST is local in nature, we are able to detect the dynamical behavior of the signal \cite{Daubechies_Lu_Wu:2011,Chen_Cheng_Wu:2014};
\item[(P3)] The signal's time-frequency representation is visually informative;
\item[(P4)] SST is ``adaptive'' to the data in the sense that the error in the estimation depends only on the first three moments of the mother wavelet instead of the profile of the mother wavelet;
\end{itemize}

In addition to the above, SST has an important property -- since the reassignment is carried out only on the frequency axis, the causality of the signal is preserved. Furthermore, with a well-chosen window, the algorithm can be implemented in real-time. Indeed, if we choose a compactly supported window function in STFT, then the analysis at time $t$ will depend only on the signal around $t$. For example, if the window function is supported on $[-\tau,\tau]$, where $\tau>0$, then the numerical computation depends on a fixed number of sample points so that SST can be implemented in real time, and the information we could obtain from SST is lagged by $\tau$. A similar argument holds for the SST based on CWT, while the lag depends on the scale; the larger the scale is, the longer the lag is. A convenient compactly supported mother wavelet based on the spline and its implementation for real-time SST could be found \cite{Chui_Lin_Wu:2014}.

Based on these properties, we are able to robustly acquire the ``phase'' of each heart beat from the ECG signal, even when the noise is large. This phase information provides a different facet about the location of R peaks. 
Thus, by integrating this information with the existing R peak detection algorithm, it is reasonable to expect an improved accuracy.

\subsection{Novel ECG morphology feature}
\label{Section:waveshapeecgqual}

The above model and discussion lead to a novel index quantifying the signal feature. 
First, extract the fundamental phase function, $\varphi_1(t)$, of the lead II ECG signal in (\ref{eq:resp_model2}).
Denote the estimated $\varphi_1$ by SST as $\tilde{\varphi}_{1}(t)$, defined within $(-\pi,\pi]$. Then we could segment the ECG signal by the estimated phase function. Precisely, denote $t_r\in\RR$ as the time stamp where $\tilde{\varphi}_{1}(nt_r)-\tilde{\varphi}_{1}((n-1)t_r)=2\pi $, with $n\in\ZZ$, $r\in\NN$. By segmenting the whole lead II ECG signal into intervals $I_r:=[t_r,t_{r+1})$, we could guarantee that precisely one R peak is inside $I_r$ with the relative phase $\zeta_r:=\tilde{\varphi}_{1}(s_r)-\tilde{\varphi}_{1}(t_r)$, where $s_r\in[t_r,t_{r+1})$ is the time stamp of the $r$-th R peak. 
Suppose a subject is healthy without any ectopic beat or arrhythmia, the R peak of each normal beat should happen at an approximately the same time inside each $I_r$; that is, $\zeta_r$ should be approximately the same for all $r$. This fact could be described in the model (\ref{eq:resp_model2}). 
On the other hand, if a subject has ectopic beat or arrhythmia, then modeling the lead II ECG signal by a single wave shape function is certainly not enough and the R peaks might have variable phases inside each interval. We can thus denote by $Y:=(\zeta_r)_{r\in\NN}$ the distribution of phases at R-peak locations and use it to characterize ECG morphology and quality. The standard deviation of $(\zeta_r)$ can provide a rough estimate of the amount arrhythmic beats. We mention that in the extremal case when a subject loss his HRV and without ectopic beats, for example a subject in the intensive care unit, then we would expect the standard deviation to be even smaller.

In addition to $Y$, we could consider the following Z Index. Denote the reconstructed $k$-th component in the model (\ref{eq:resp_model2}) as $\tilde{B}_{\II,l}(t)\cos(2\pi \tilde{\varphi}_l(t))$. Define the following index for the $i$-th beat
\[
Z_i=\left[\tilde{B}_{\II,1}(t)\ud t,\ldots, \tilde{B}_{\II,D}(t)\ud t\right]^T\in\RR^D.
\]
Note that when the wave shape function is fixed, $Z_i$ depends on Fourier coefficients of the wave shape function and the amplitude modulation function. In (\ref{eq:resp_model2}), it would characterize the time-varying shape function, which allows us a better quantification of the ectopic or arrhythmic beats.

\section{R-peak detection algorithm}\label{Section:RpeakAlgo}
We investigated several R-peak detection algorithms that were reported to perform well and selected a few for which algorithm details were well described. Algorithms based on \cite{Pan_Tompkins:1985,Hamilton_Tompkins:1986,Kathirvel2011,Elgendi2013} were implemented. For testing, we used the MIT-BIH ECG arrhythmia database which consists of 48 annotated records containing ECG MLII lead and any one of V1, V2, V4 or V5 precordial lead recording \cite{mitbih}. After optimization of the parameters for each algorithm according to each respective manuscript, the best performing algorithm in our tests was that proposed by Elgendi \cite{Elgendi2013}. We briefly describe the steps involved hereafter. 
First the ECG signal is bandpass filtered with a third order Butterworth filter with a passband between 8Hz and 20Hz to remove high-frequency noise and baseline wander. The filtered signal is then squared to enhance peaks. The rough locations of QRS complexes are found by comparing the outputs from two moving average filters ($MA_\QRS$ and $MA_\beat$), whose window sizes are chosen such that they match approximately the duration of a typical QRS complex (deemed $W_1$) and heartbeat (deemed $W_2$), respectively. The final blocks of interest containing potential R-peaks are found by thresholding $MA_\QRS$, using an adaptive threshold derived from $MA_\beat$ and the mean of the squared filtered ECG signal. Finally, R-peak locations are found by looking at the maximum absolute value within each block of interest of duration larger than $W_1$.

An additional stage not contained in the original paper was added to reduce the incidence of false R-peaks detected too close to each other, using a refractory period of $250$ms, similar to that found in for e.g. \cite{Pan_Tompkins:1985}. Whenever two peaks are detected within less than 250ms of each other, the peak with the sharpest slope is retained.

\section{Our proposed algorithm}\label{Section:OurAlgorithm}
The R-peak detection described in section \ref{Section:RpeakAlgo} provided a list of potential R-peaks. The synchrosqueezing transform was applied to signal $X_m(t)$ derived from the ECG waveform and defined as follows:
\begin{equation}
X_m(t) = \log \left (1+\left |(E_\II(t)-T(t))\right | \right )-\mu
\end{equation} 
Where $E_\II(t)$ is the ECG signal, $T(t)$ a lowpass filtered version of $E_\II(t)$ and $\mu$ is the mean of $\log \left (1+\left |(E_\II(t)-T(t))\right | \right )$.   

For computational tractability,  $X_m(t)$ was divided in blocks of $2^{12}$ samples and to minimize edge effects, the blocks were overlapped by 50\%. For each block, a wavelet-based SST was performed, using a mother wavelet $\psi$ so that
\begin{equation}
\hat{\psi}(\xi)=\exp\left(\frac{1}{\left(\frac{\xi-1}{\sigma}\right)^2-1}\right)\chi_{[1-\sigma,1+\sigma]},
\end{equation}
where $\sigma$ means the Full Width at Half Maximum (FWHM) and is chosen as $0.8\times IFF$, where $IFF$ is the estimated instantaneous frequency derived from the R-peak detected over each block. The center of the mother wavelet was chosen to be equal to $IFF$ as well. The instantaneous phase was estimated by tracking the support of the dominant curve in the synchrosqueezed transform within a frequency band of $\pm 0.3 \times IFF$. 

Following the discussion from \ref{Section:waveshapeecgqual} and using the same notations, we expected to find an R-peak within each interval $I_r:=[t_r,t_{r+1})$ representing successive phase shifts (from $\pi$ to $-\pi$) of the resulting instantaneous phase $\tilde{\varphi}_{1}(t)$. In order to recover missed beats, a search was performed on the phase shift intervals where no R-peak was detected. For each $r$, if $T_{r+1}-T_{r}>200ms$ and if $P=\max_{t\in\left[T_{r}, T_{r+1} \right]} (E_\II(t))>TH1$ and if $|\tilde{\phi}(t=t_P)-r|<\pi /2$, where $t_P$ is the time corresponding to the potential R-peak $P$, then $P$ is likely an R-peak. Threshold $TH1$ was defined as a scaling factor (set as 0.7) times the difference between the median of the last 4 peaks and the median of the ECG signal over the same period. 

\section{Heart beat classification}
\label{sec:heartbeatclass}

Following the AAMI standard \cite{aami}, we grouped the different beat types from the MIT-BIH database into five classes. Class N contains normal and bundle branch block beat types; Class S contains supraventricular ectopic beats, class V contains ventricular ectopic beats, class F contains fusion of normal and ventricular ectopic beats and class Q contains paced beats, fusion of paced and normal beats and unknown beat types. According to common recommended practice, we excluded the four recordings containing paced beats. In addition, we did not consider the class Q in any of the following beat classification tasks.

Furthermore, in order to compare with algorithms from the recent literature \cite{Dechazal2004,llamedo2011heartbeat,ye2012heartbeat}, we separated the remaining 44 recordings into a training set (DS1) and a testing set (DS2), each containing 22 recordings with the same proportion of beat types. A summary of the beat types, classes and datasets is given in Table \ref{Tab:beattypes}. 

\begin{table}[ht]
\caption{Heartbeat types from the full database and dataset 1 and 2 from the MIT-BIH Arrhythmia database. N: normal; L: Left bundle branch block; R: Right bundle branch block; A: Atrial premature; a: aberrated atrial premature; J: Nodal (junctional) premature; S: Supraventricular premature or ectopic; V: Premature ventricular contraction; F:Fusion of ventricular and normal; e: Atrial escape; j: Nodal (junctional) escape; E: Ventricular escape; Q: Unclassifiable beat}
\centering 
\resizebox{\textwidth}{!}{\begin{tabular}{c c c c c c c c c c c c c c}
\hline\hline \noalign{\smallskip}
Heartbeat type& N& L& R& A& a& J& S& V& F& e& j& E& Q\\
Heartbeat class& N& N& N& S& S& S& S& V& F& N& N& V& Q\\ [0.5ex]
\hline \noalign{\smallskip}
All records	& 74546 & 8075& 7259& 2546& 150& 83& 2& 3903& 803& 16& 229& 106& 15\\
DS1& 38102& 3949& 3783& 810& 100& 32& 2& 3683& 415& 16& 16& 105& 8\\
DS2& 36444& 4126& 3476& 1736& 50& 51& 0& 3220& 388& 0& 213& 1& 7\\
 \hline
\end{tabular}}
\label{Tab:beattypes}
\end{table}

Support vector machines (SVM) have been reported to perform well for classifying beat types \cite{ye2012heartbeat,zhang2014heartbeat} and were selected in this work. We trained SVM classifiers using  a set of simple features derived from either one of the two ECG leads: 
\begin{itemize}
\item SST-derived phase at R-peak locations
\item R-peak amplitude
\item time difference between current and previous beat (at R-peak)
\item time difference between current and next beat (at R-peak)
\item Average R-peak to R-peak interval over 10 beats
\item Duration of the QRS complex.
\end{itemize}
These additional features besides the SST-derived features were chosen because they are easily derived from the ECG, they are relatively robust in the presence of noise, they are linked to characteristic aspects of different beat types and they have been shown to give good performance \cite{Dechazal2004,llamedo2011heartbeat,ye2012heartbeat}.
The features derived at the true R-peak locations available from the MIT-BIH database were used for the heart beat classification in order to have a comparable number of beats as found in the existing literature. We used the $C$-support vector classification formulation with a radial basis function (RBF) kernel and a weighted $C$, where each class weight is optimally selected to penalize the class according to its prevalence \cite{Vapnik1998}. The one-vs-one approach was used to train the multiclass SVM classifier, using the libsvm library \cite{CC01a}. Both constraint $C$, RBF kernel parameter $\gamma$ and weights were optimized on a 10-fold cross-validation on the training data set (DS1). Final performance results were evaluated on the testing set (DS2). Two sets of SVM classifiers were investigated. We first checked that beats originating from the sinus nodes (class N) could be separated from all other classes (S,V,Q and F). A second SVM classifier was trained to separate between nodal, supraventricular ectopic, ventricular ectopic, and fusion beat groups (N, S, V, F, respectively). 

\section{Quality estimation}

The distribution of phase at R-peak locations was also compared to an estimate of the level of noise, to see if it could be used as an additional index of quality of the ECG signal. The reference quality level was determined by thresholding the signal quality index (SQI) derived from the ECG waveform (see \cite{clifford_signal_2012} for details of the SQI) in three levels: poor, medium and high quality, based on the SQI thresholds of 33\% and 66\%. Using the phase, amplitude at R-peak location and R-R intervals, we trained a simple K-nearest neighbor (KNN) classifier with $K=9$ on a combination of datasets DS1 and DS2. The resulting dataset had 100689 beats, with a majority of high-quality beats (86.21\%).  We estimated the performance of the KNN  using a 10-fold stratified cross-validation where each fold contained an equivalent proportion of poor, medium and high-quality beats randomly selected from all recordings. 

\section{Results}\label{Section:Result}
Table \ref{Tab:elgendisstcwtmitbih} shows the beat detection results for both leads and a slight reduction in the number of missed beats and false beats when relying on the instantaneous phase derived from the SST to recover missed beats. Detection results were lower on the second ECG channel in each record, which consisted of a mix of precordial leads (V1, V2, V4 or V5). 

\begin{table}[ht]
\caption{R-peak detection results on the MIT-BIH database. The total number of beats considered was 109494 over 48 records. TP: number of true beats; FN: number of missed beats; FP: number of false beats; Se: Sensitivity; +P: positive predictive value.}
\centering 
\resizebox{\textwidth}{!}{\begin{tabular}{c c c c c c c}
\hline\hline \noalign{\smallskip}
Method & TP & FN & FP & Se (\%)& +P (\%) \\ [0.5ex] 
\hline \noalign{\smallskip}
\cite{Elgendi2013} + SST-CWT (lead II) & 109299 & 195 & 128 & 99.82\% & 99.88\%\\
\cite{Elgendi2013} (lead II) & 109289 & 205 & 133 & 99.81\% & 99.88\%\\
\cite{Elgendi2013} + SST-CWT (precordial leads)  & 104743 & 3784 & 4756 & 95.66\% & 96.51\%\\
\cite{Elgendi2013} (precordial leads)  & 104475 & 3661 & 5022 & 95.41\% & 96.61\%\\
 \hline
\end{tabular}}
\label{Tab:elgendisstcwtmitbih}
\end{table}

\begin{figure}[tbp]
\centering
\includegraphics[clip=true, width=0.9\textwidth]{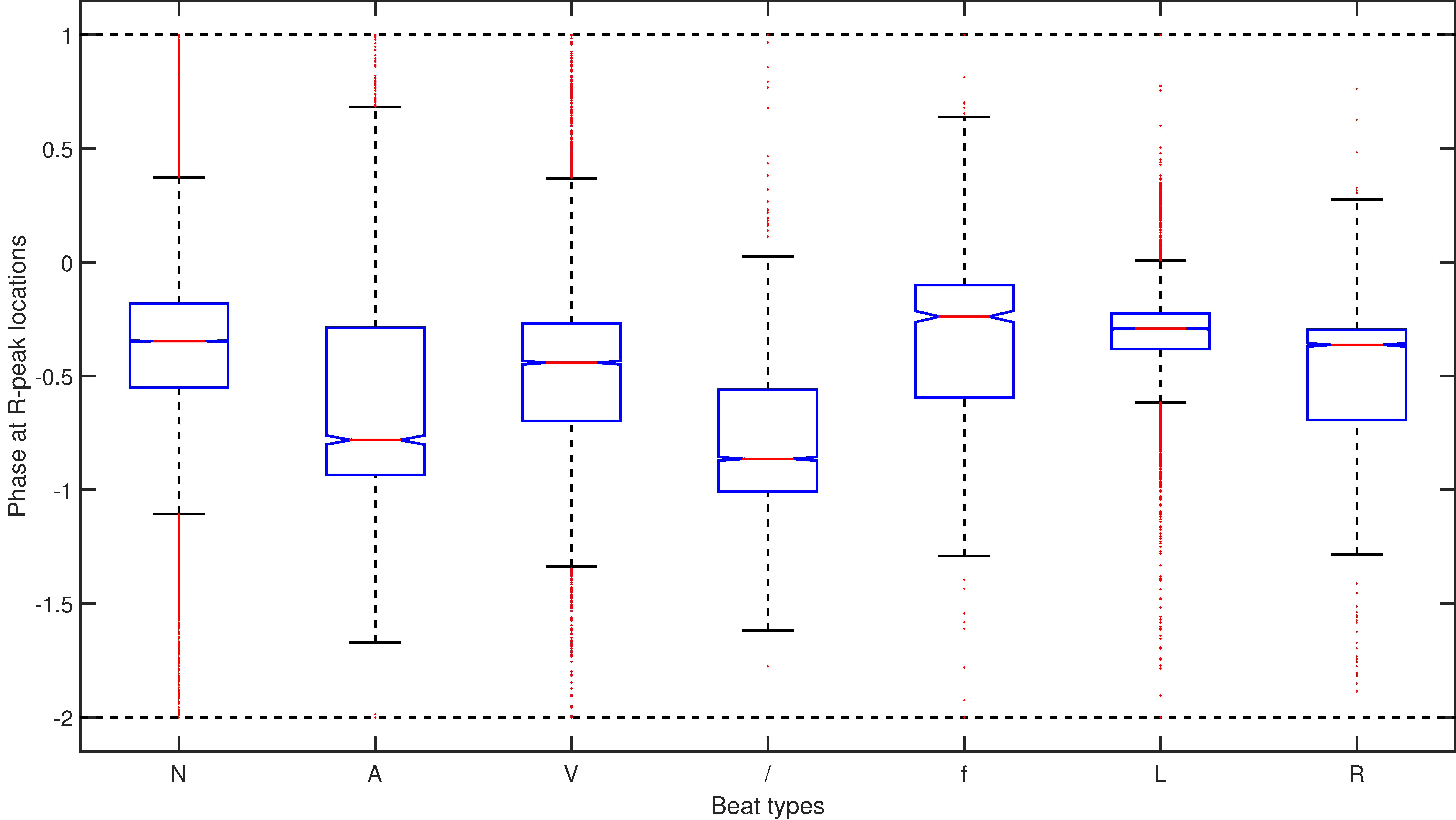}
\caption{Boxplots of the distributions of phase at R-peak locations for the main beat types present in the MIT-BIH database. N: normal (75052 beats); A: Atrial premature (2546 beats); V: Premature ventricular contraction (7130 beats); /: paced (7028 beats); f: Fusion of paced and normal (982 beats); L: Left bundle branch block (8075 beats); R: right bundle branch block (7259 beats).}
\label{fig:beattypevsphase}
\end{figure}

Perhaps more interesting was the distribution of the phase, especially at R-peak locations. First, we looked at the difference in phase between the different beat labels, as shown in Fig. \ref{fig:beattypevsphase}. There was a  significant difference between the median phase of each beat type, suggesting that it could be used as a feature for classifying beat types.

The first SVM classifier was trained to differentiate between beats originating from the sinus node vs ectopic and fusion beats. The features used are the six features described in section \ref{sec:heartbeatclass}, derived from the ECG lead II only. Optimal parameters for the weighted SVM formulation described in Section \ref{sec:heartbeatclass} were found to be $C=8.38$ and $\gamma=0.52$, with a weight of 1 for class N and 5 for class S/V/F/Q, based on a 10-fold cross-validation on training dataset DS1. Sensitivity, specificity and accuracy of the trained classifier assessed on test set DS2 were 92.12\%, 86.69\% and 87.29\%, respectively (when considering the abnormal class as the positive class). The corresponding confusion matrix is shown in Table \ref{Tab:svm2class}. The total number of beats from DS2 used in the classification was 49603, since not all features could be derived from each beat, mostly the first few and last beats of each recording.

\begin{table}[ht]
\caption{Confusion matrix summarizing results from applying a binary SVM classifier on dataset DS2, to classify beats originating from sinus nodes vs other beats. N: sinus node beats; S/V/F/Q: All non sinus node beats}
\centering 
\begin{tabular}{c c c c c c}
\hline\hline \noalign{\smallskip}
& & \multicolumn{2}{c}{Predicted}\\
& & S/V/F/Q & N \\ [0.5ex] 
\hline \noalign{\smallskip}
 {\multirow{2}{*}{\rotatebox[origin=c]{90}{Ref.}}}&S/V/F/Q & 5004 & 428 \\
& N  & 5879  &  38292\\
 \hline
\end{tabular}
\label{Tab:svm2class}
\end{table}

A second SVM classifier with parameters $C=3.98$, $\gamma =1.98$, and weights $w_1=0.42$, $w_2=55$, $w_3=0.85$ and $w_4=5.3$ for classes 1 to 4, respectively, was found to be optimal on a 10-fold cross-validation on the training dataset (DS1), for differentiating between the four main AAMI classes (N,S,V and F). The performance of this classifier was assessed on the test set DS2 and the resulting confusion matrix is reported in Table \ref{Tab:svmcfmat}. The overall accuracy is 83.20\% and the per class sensitivity (or recall) and positive predictive value (or precision) are reported in Table \ref{Tab:svmds2results}, along with the performance of selected methods from the literature. Per class performance measures are defined as follows: Denote $n_{i,j}$ to be the $(i,j)$-th entry of the confusion matrix. The {\it sensitivity} of the class $i$ prediction is defined as $Se(i):=\frac{ n_{i,i}}{\sum_{k}n_{i,k}}$; the positive predictive value of the class $i$ denoted as $+P(i):=\frac{ n_{i,i}}{\sum_{k}n_{k,i}}$. The overall accuracy (ACC) denoted as $ACC:=\frac{\sum_{i=1}^{5}n_{i,i}}{\sum_{i,j=1}^{5}n_{i,j}}$. 
Note that the other methods shown in table \ref{Tab:svmds2results} all use 2 ECG leads and many more features.
A separate SVM classifier was also optimized on the same features derived from the alternate ECG channel from the MIT-BIH database records. All performance results are shown in Table \ref{Tab:svmds2results}. Finally, in order to compare with the multilead systems from Table \ref{Tab:svmds2results}, we also show the results from training a classifier using the merged decision from two optimized SVM classifiers trained using the features described in section \ref{sec:heartbeatclass}, extracted from both channels of the MIT-BIH records. The merged decision was the most common beat class from the one-vs-one (class) decision functions.

\begin{table}[ht]
\caption{Confusion matrix summarizing results from applying the proposed SVM classifier (using lead II features) on dataset DS2, to classify AAMI beat types. N: beats originating from sinus node; S: Supraventricular ectopic; V: Ventricular ectopic; F: fusion of normal and ectopic.}
\centering 
\begin{tabular}{c c c c c c}
\hline\hline \noalign{\smallskip}
& & \multicolumn{4}{c}{Predicted}\\
& & N & S & V & F \\ [0.5ex] 
\hline \noalign{\smallskip}
 {\multirow{4}{*}{\rotatebox[origin=c]{90}{Reference}}}&N & 36721 & 2694 & 553 & 4203 \\
& S  & 239  & 1489 & 81 & 26 \\
& V  & 124  & 473 & 2487 & 125  \\
& F & 33 & 7 & 25 & 323 \\
 \hline
\end{tabular}
\label{Tab:svmcfmat}
\end{table}

\begin{table*}[t]
\caption{Beat type classification performance of the proposed SVM classifier and feature set, on dataset DS2, and comparison with other methods from the literature: 1: \cite{llamedo2011heartbeat} 2:\cite{zhang2014heartbeat}, 3: \cite{ye2012heartbeat}, 4: \cite{Dechazal2004}, 5: proposed SST-based method (with lead MLII). 6: proposed SST-based method (with lead V1, V2, V4 or V5). 7: proposed SST-based method (both leads, merged decision from each lead). Se: Sensitivity; +P: Positive predictive value, ACC: accuracy. \# feat: number of features extracted from the ECG. \# leads: number of ECG leads used for the classification; $\dagger$: the final classifier uses 18 PCA features derived from the 132 initial ones.}
\centering 
\resizebox{\textwidth}{!}{\begin{tabular}{c c c c c c c c c c c}
\hline\hline \noalign{\smallskip}
 & \# features / &  \multicolumn{2}{c}{N} & \multicolumn{2}{c}{S} & \multicolumn{2}{c}{V} & \multicolumn{2}{c}{F} & Tot.\\
 & \# leads & Se (\%) & +P (\%) & Se (\%) & +P (\%) & Se (\%) & +P (\%) & Se (\%) & +P (\%) & ACC (\%)\\ [0.5ex] 
\hline \noalign{\smallskip}
1 & 39 / 2 & 77.55 & 99.47 & 76.46 & 41.34 & 82.94 & 87.97 & 95.36 & 4.23 & 78.00 \\
2 & 46 / 2 & 88.94 & 98.98 & 79.06 & 35.98 & 85.48 & 92.75 & 93.81 & 13.73 & 88.34\\
3 & $132^\dagger$ / 2& 88.51 & 97.54 & 60.80 & 52.34 & 81.49 & 61.38 & 19.59 & 2.50 & 86.40\\
4 & 52 / 2& 86.86 & 99.16 & 75.90 & 38.50 & 77.70 & 81.90 & 89.43 & 0.08 & 81.90 \\
5 & 6 / 1& 83.13 & 98.93 & 81.14 & 31.93 & 77.50 & 79.05 & 83.25 & 6.91 & 82.70 \\
6 & 6 / 1& 85.34 & 98.50 & 62.72 & 37.15 & 79.56 & 62.68 & 86.60 & 8.07 & 84.13 \\
7 & 12 / 2& 85.25& 99.18&81.85 &39.20 &79.00 &70.75 &85.05 &7.83 & 84.71 \\
 \hline
\end{tabular}}
\label{Tab:svmds2results}
\end{table*}

Results from the classification of ECG quality levels based on the phase, amplitude and R-R interval are reported in Table \ref{Tab:KNNquality}. 
\begin{table}[ht]
\caption{Confusion matrix and performance statistics for the 10-fold, cross-validated performance of the KNN classifier (9 neighbors) using 100689 beats from the MIT-BIH database. LQ: Low quality, MQ: medium quality; HQ: high quality; Se: Sensitivity; +P: positive predictive value; Overall accuracy is 93.91\%}
\centering 
\begin{tabular}{c c c c c c}
\hline\hline \noalign{\smallskip}
 & LQ & MQ & HQ & Se (\%) & +P (\%)\\ [0.5ex] 
\hline \noalign{\smallskip}
LQ & 6322 & 42 & 1910 & 76.41 & 93.66  \\
MQ  & 75  & 2406 & 3128 & 42.90 & 78.45  \\
HQ  & 355  & 620 & 85831 & 98.88 & 94.46  \\
 \hline
\end{tabular}
\label{Tab:KNNquality}
\end{table}

\begin{figure}[tbp]
\centering
\subfigure[]{\includegraphics[clip=true, width=0.7\textwidth]{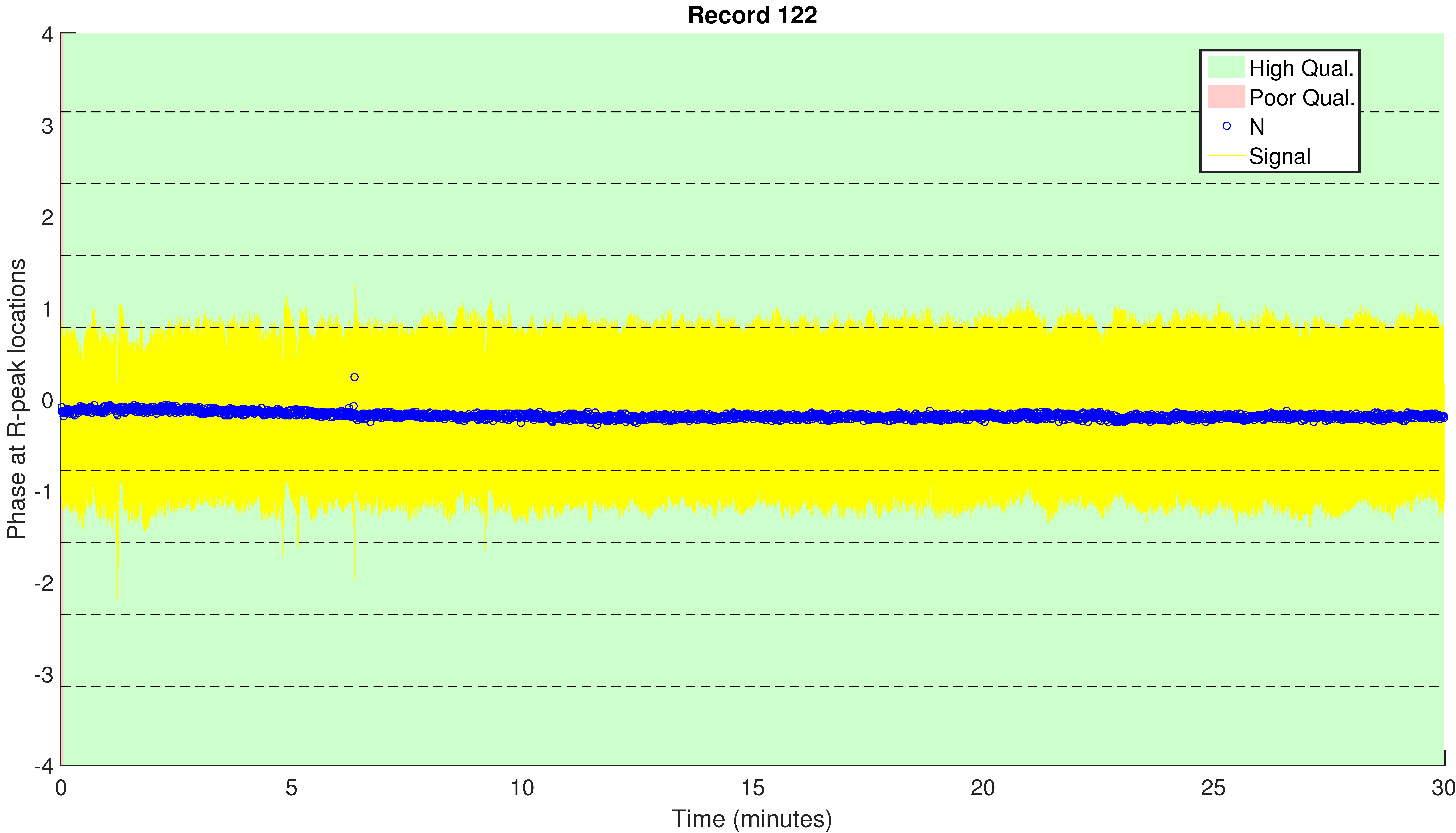}}
\subfigure[]{\includegraphics[clip=true, width=0.7\textwidth]{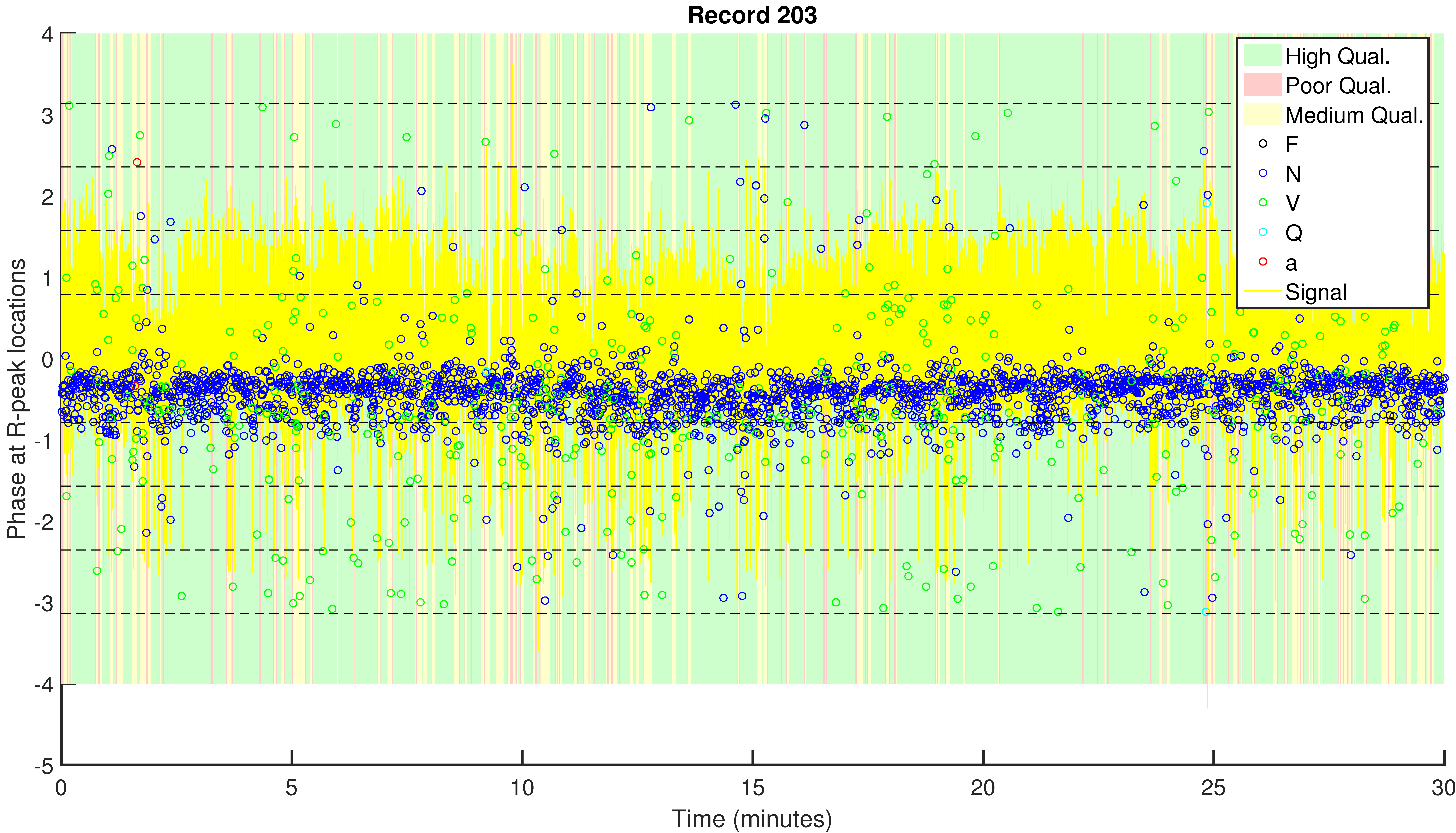}}
\subfigure[]{\includegraphics[clip=true, width=0.7\textwidth]{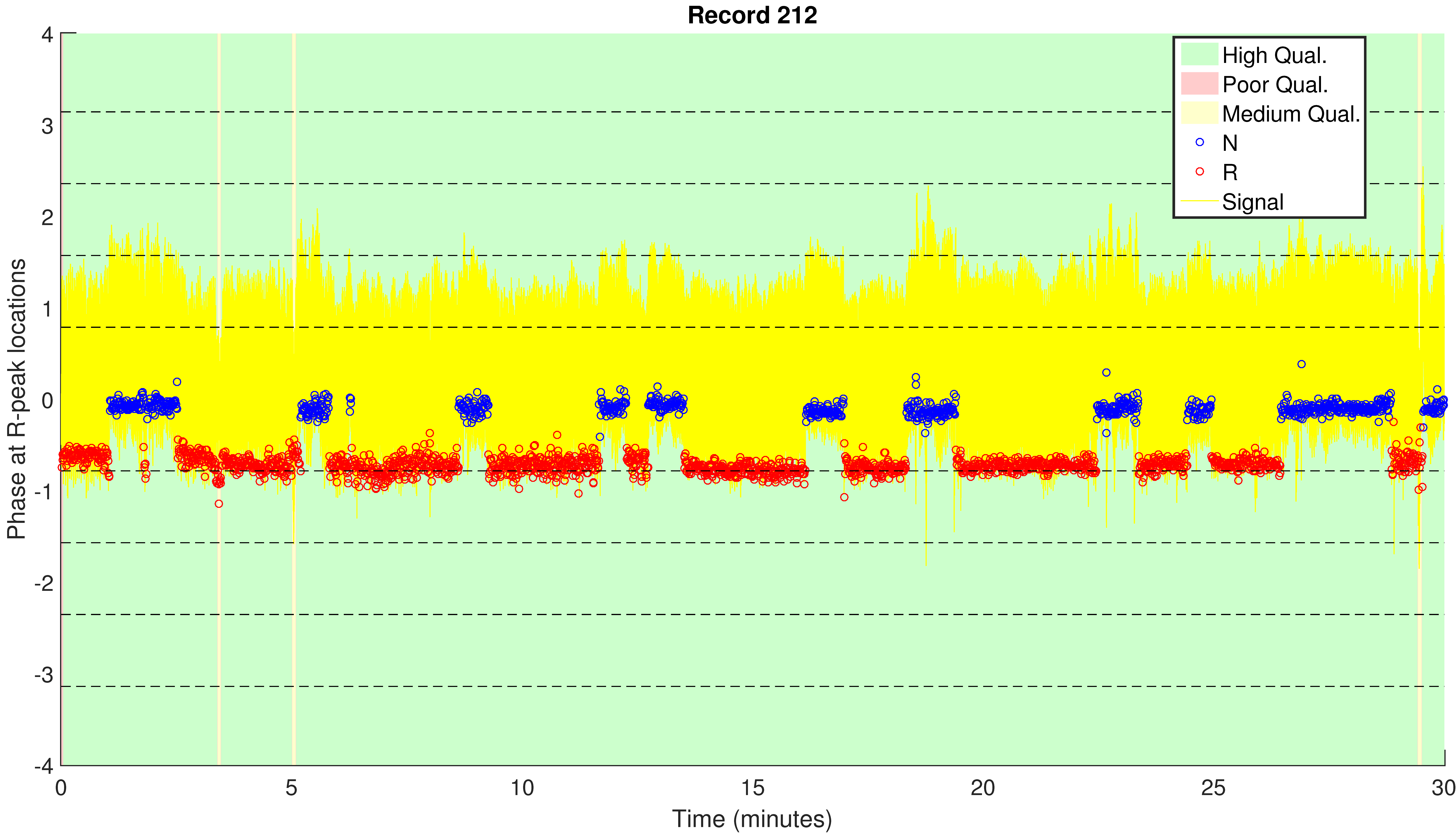}}
\caption{Examples of different phase patterns for: (a) a clean signal with only normal sinus rhythm beats; (b) a noisy signal with arrhythmic beats; (c) a distinctive alternative phase pattern between normal and right branch block beats.}
\label{fig:phasedispersionexamples}
\end{figure}

Finally, Fig. \ref{fig:phasedispersionexamples} shows three typical examples of the difference in distribution of the phase at R-peak locations for three records from the MIT-BIH database. The spread of the distribution of phase at R-peak locations for a clean signal comprised of normal sinus rhythm beats is minimal, whereas it is very large for the noisy signal with many arrhythmic beats. A sudden shift in the R-peak phase can also indicate a switch between normal and abnormal beats as shown in Fig. \ref{fig:phasedispersionexamples} c). 

\section{Discussion}\label{Section:Discussion}

The addition of the synchrosqueezing transform to extract the instantaneous phase information from the ECG waveform marginally improves on the R-peak detection of currently available algorithms. However, recovering the phase information at R-peak location shows intriguing potential for characterizing the beats extracted, as well as complementing the assessment of the quality of the ECG waveform and R-peak time series. Results from section \ref{Section:Result} indicate that a simple combination of phase and amplitude at R-peak locations from a single ECG channel is sufficient to reach reasonable classification performance for differentiating between normal and abnormal beats (ACC: 96.24\%, Se: 98.50\%, +P: 97.31\%). 

The heartbeat classification using 6 simple features derived from a single ECG channel, including the SST-derived estimation of the phase at R-peak location, compared well with recent methods (cf. Table \ref{Tab:svmds2results}). This is significant if we consider that the compared methods use anywhere between 39 and 152 features derived from 2 ECG leads. When using only lead II, the lowest sensitivity obtained by our classifier was 77.5\% for the ventricular ectopic class, while for all other beat classes the sensitivity was over 80\%. It is important to point out that in our weighted SVM formulation, the misclassification costs associated with each beat class could be optimized differently, for instance to increase the classification performance of a particular beat class. Similar performance was obtained when using features derived from the second ECG channel from each record, although performance in ectopic beat classes was slightly degraded, particularly for the supraventricular ectopic class. Using our set of features extracted from both leads yield yet improved performance, with more balanced results across all beat types.

The limited number of morphological features used in our classifier, in particular the lack of characterization of the P- and T-wave may have contributed to the misclassifition of supraventricular ectopic beats as normal beats and ventricular ectopic beats as supraventricular ectopic beats. Perhaps another factor is the fact that the separation into DS1 and DS2 used in the literature may not be ideal for certain beat classes, as noted by Llamedo et al. \cite{llamedo2011heartbeat}. Indeed, a single record in the test set DS2 contains over 50\% of the atrial premature beats, which are the main beat types in the supraventricular ectopic group. Testing our methodology on a larger dataset of mixed annotated databases of ECG recordings could help further elucidate the difference in performance between groups of beat types.
Nonetheless, with the wide availability of small, compact, single lead ambulatory ECG monitoring devices and commerical mobile health devices, we believe there is much potential in using single lead features, in particular the phase information, for fast heartbeat classification. These devices often have non-standard lead placement and can be significanlty noisy, which would seriously affect the extraction of many of the morphological features proposed in the current literature on heartbeat classification. We believe that our classifier based on simple features would be most beneficial in these cases. Another application of our technique is in fetal ECG recordings, where single-lead noisy ECG signals are common and could benefit from improved heartbeat detection and classification.  

The phase information at R-peak locations was shown to be tightly linked to the quality of the underlying ECG signal as well. High and low quality beats were well characterized by the instantaneous phases defined by the phase and amplitude features. The separation between medium and high quality beats was more difficult to discern. This is expected as the SQI threshold to separate between medium and high quality beats was arbitrarily set to 66\% and the estimated quality of many beats were close to that threshold.

Finally, in this work we restricted the SST-derived instantaneous phase information to the first oscillatory component ($D=1$) in \ref{eq:resp_model2}. We anticipate that the use of multiple oscillatory components $D>1$, corresponding to a time-varying shape function, would improve the local estimation of the ECG signal. Alternatively, using features directly from the time-varying power spectrum (tvPS) could provide additional benefit in discriminating between beat types, at the cost of increased computational requirements. For instance, we could use an {\em Optimal Transport} (OT) distance (also called the Earth Mover distance) between time-varying spectra of duration sufficient to capture information of the ventricular activity of each beat and use a clustering algorithm to recover the beat classes.

\section{Conclusions}
In this manuscript, we characterized the ECG signal as an adaptive non-harmonic model and used the synchrosqueezing transform and  to derive characteristics of the ECG signal, in particular the instantaneous phase. We demonstrated that using the SST-derived phase and amplitude of the ECG signal at R-peak locations, along with additional QRS timing features, can be used to classify non-life-threatening arrhythmias and provide an additional index of quality of the underlying ECG signal. Despite using only one lead and 6 simple features, the heart beat classification performance on the MIT-BIH database was similar to methods using multilead ECG and a larger number of features. Adding key morphological features (e.g. P-wave characteristics), would likely increase the performance further, particularly for supraventricular beats. These results merit further testing on another diverse annotated databases to better assess generalizability.

\section{Acknoledgement}
Hau-tieng Wu acknowledges the support of Sloan Research Fellow FR-2015-65363. Christophe Herry is a patent holder related to waveform quality assessment necessary for variability analysis. 
We thank the reviewers for their most helpful comments and suggestions.

\bibliographystyle{amsplain}
\bibliography{SST_Rpeak_PM_R3}
\end{document}